\documentclass[sigconf]{acmart}

\AtBeginDocument{%
  }

\setcopyright{acmlicensed}
\copyrightyear{2026}
\acmYear{2026}
\acmDOI{XXXXXXX.XXXXXXX}
\acmConference[Conference acronym 'XX]{Make sure to enter the correct
  conference title from your rights confirmation email}{June 03--05,
  2026}{Woodstock, NY}
\acmISBN{978-1-4503-XXXX-X/2026/06}

\usepackage{graphicx}
\usepackage{float}
\usepackage[normalem]{ulem}
\usepackage{array}
\usepackage{url}
\usepackage{tikz}
\usepackage{colortbl}
\usepackage{eucal}
\usepackage{bm}
\usepackage{multirow}
\usepackage{enumitem}
\usepackage{xspace}
\usepackage{amsfonts}
\usepackage{booktabs}
\usepackage[english]{babel}
\usepackage{blindtext}
\usepackage{pifont} 
\usepackage{stackengine}
\usepackage{setspace}
\usepackage{siunitx}
\usepackage{threeparttable}
\usepackage{tabularx}
\usepackage{caption}
\usepackage{subcaption}
\usepackage{hyperref}
\usepackage{wrapfig}

\usepackage{epstopdf}

\usepackage{algorithm}
\usepackage{algpseudocode}
\usepackage{amsmath}

\newcolumntype{C}[1]{>{\centering\arraybackslash}p{#1}}
\newcolumntype{L}[1]{>{\raggedright\arraybackslash}p{#1}}

\makeatletter
\def\@authorsaddresses{}
\makeatother

\def\ourSystem{\texttt{FM-CAC}\xspace}

\def\ie{\textit{i.e.},\xspace}
\def\eg{\textit{e.g.},\xspace}

\def\vs{\textit{vs.}\xspace}

\newlength{\vslen}
\newlength{\vslentw}
\definecolor{award}{HTML}{C00000}

\usepackage{xcolor}

\definecolor{deepblue}{HTML}{2B00A8}

\usepackage{listings}
\usepackage{xcolor}

\definecolor{codegreen}{RGB}{0,128,0}
\definecolor{codepink}{RGB}{199,37,78}
\definecolor{codeblue}{RGB}{0,0,255}
\definecolor{darkblue}{RGB}{0,0,180}

\begin{document}

\title[\ourSystem{}: Carbon-Aware Control for Battery-Buffered Edge AI via Time-Series Foundation Models]{\ourSystem{}: Carbon-Aware Control for Battery-Buffered Edge AI \\ via Time-Series Foundation Models}

\author{Kang Yang}
\orcid{0000-0001-8248-4894}
\affiliation{\institution{University of California, Los Angeles}
\city{Los Angeles}
\country{USA}}
\email{kyang73@g.ucla.edu}

\author{Walid A. Hanafy}
\orcid{0000-0001-5765-8194}
\affiliation{\institution{University of Massachusetts Amherst}
\city{Amherst}
\country{USA}}
\email{whanafy@cs.umass.edu}

\author{Prashant Shenoy}
\orcid{0000-0002-5435-1901}
\affiliation{\institution{University of Massachusetts Amherst}
\city{Amherst}
\country{USA}}
\email{shenoy@cs.umass.edu}

\author{Mani Srivastava}
\orcid{0000-0002-3782-9192}
\affiliation{\institution{University of California, Los Angeles}
\city{Los Angeles}
\country{USA}}
\email{mbs@ucla.edu}

\begin{abstract}

As edge~AI deployments scale to billions of devices running always-on, real-time compound AI pipelines, they represent a massive and largely unmanaged source of energy consumption and carbon emissions.
To reduce carbon emissions while maximizing Quality-of-Service~(QoS), this paper proposes \ourSystem{}, a proactive carbon-aware control framework that leverages a battery as an active temporal buffer.
By decoupling energy acquisition from energy consumption, \ourSystem{} can maximize the use of low-carbon energy, substantially reducing carbon emissions.
At each control step, \ourSystem{} jointly optimizes the software pipeline variant, the hardware operating point, and the battery charging and discharging actions.
To support this decision process, \ourSystem{} leverages edge-friendly~Time-Series Foundation Models~(TSFMs) for zero-shot carbon forecasting and integrates these forecasts into a dynamic programming solver with deferred cost attribution to prevent myopic battery depletion.
Results show that \ourSystem{} reduces carbon emissions by up to~65.6\% while maintaining near-maximum inference accuracy.

\end{abstract}




\maketitle

\section{Introduction}
\label{sec:introduction}

While economic and regulatory incentives have driven emission reductions in datacenters, the rapidly growing edge remains a major and largely unmanaged source of carbon emissions, operating continuously with little or no carbon awareness~\cite{radovanovic2022carbon}.
Its scale is substantial:~Internet of Things~(IoT) devices are projected to reach~39~billion by~2030~\cite{iotanalytics2025}, smartphones to exceed~8~billion~\cite{bankmycell}, and surveillance cameras to surpass~1~billion deployments~\cite{ihs2019cameras}.
Consequently, even modest per-device reductions can yield significant aggregate benefits. 
For example, reducing emissions by~3\,gCO$_2$ per camera per day would avoid~1.1~million metric tons of CO$_2$ annually, equivalent to the emissions of~240{,}000 cars~\cite{gas_car}.

\begin{figure}[t]
    \centering
    \includegraphics[width=0.99\linewidth]{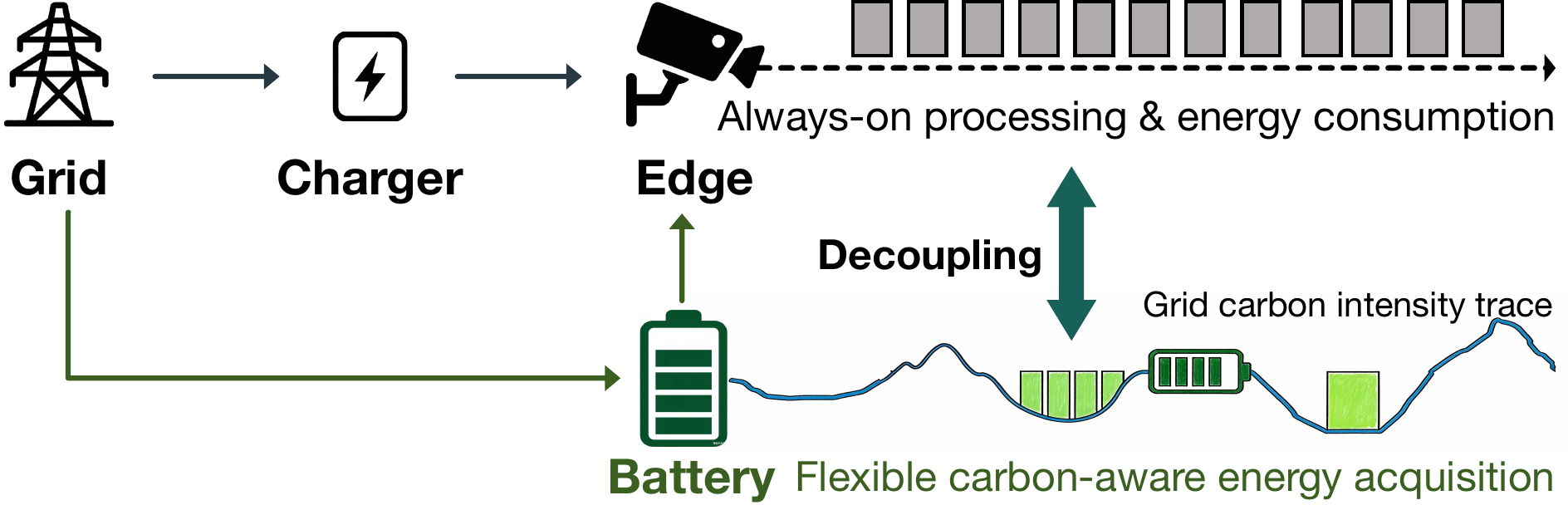}
    \caption{Illustration of battery-buffered edge AI. While edge systems draw directly from the \emph{GRID} for always-on inference, \ourSystem{} treats the \emph{BATTERY} as an active temporal buffer.}
     \Description[]{}
    \label{fig:scenario}
\end{figure}

As illustrated in Figure~\ref{fig:scenario}, edge devices such as security cameras~\cite{kim2024statues, nest_cam_floodlight} continuously capture and process data in real time, running compound~AI services~\cite{compound-ai-blog}, \eg~multi-stage pipelines for trigger, detection, tracking, and refinement. 
Existing edge systems typically power these always-on services directly from the grid, leaving their carbon footprint exposed to temporal variation in grid carbon intensity. 
Meanwhile, many edge devices already include embedded batteries for operational purposes, such as outage backup or solar-powered deployment~\cite{hasan2023battery, nest_cam_floodlight, tektelic_kona_micro}.
In this paper, we repurpose these batteries from passive fail-safes into \emph{active temporal buffers} that decouple \emph{energy acquisition} from \emph{energy consumption}. 
This enables charging during low-carbon periods and discharging during high-carbon periods, providing a practical mechanism for proactive emissions reduction without interrupting always-on inference.

However, realizing such active temporal buffering introduces a challenging control problem. 
To minimize carbon emissions while preserving Quality-of-Service~(QoS) for compound~AI pipelines, the system must jointly orchestrate \textbf{three tightly coupled control dimensions}:
\emph{1)~software configuration},~\eg pipeline variant selection, \emph{2)~hardware state},~\eg CPU/GPU frequency scaling, and \emph{3)~battery charging policy}. 
These decisions directly couple instantaneous service quality with future energy availability. 
For example, selecting a high-accuracy pipeline can accelerate battery depletion and force later grid draw during high-carbon periods to avoid service disruption.
Conversely, a conservative hardware mode may lower carbon emissions but unnecessarily sacrifice inference accuracy during low-carbon windows.
This intricate trade-off, complicated by stochastic grid carbon intensity, demands proactive control.

While carbon or energy optimization have been studied in edge systems~\cite{bullo2024energy, zhang2025e4, jeon2023harvnet, han2025dvfs, dai2024energy, siar2025energy, ma2025greening, zhu2025does, hsu2006adaptive, cho2024clover, xu2018direct, Liang2023:Delen,park2023energy}, datacenters~\cite{wiesner2021:letswait, maji2023multicloud, jiang2024ecolife, kim2023greenscale, su2023learning, lechowicz2025online}, and electric vehicle~(EV) charging~\cite{faruqui2011smartprices, jones2022tou, cheng2022carbonawareev, dixon2020evcharging, li2023mef_tou}, these paradigms remain mismatched to always-on edge~AI. 
Existing methods either optimize energy efficiency alone or reduce emissions through spatiotemporal workload shifting and long idle windows.
In contrast, edge inference permits neither workload deferral nor spatial flexibility, yet it retains intrinsic elasticity across three control dimensions. 
This requires a different form of carbon-aware control under~QoS constraints.

To jointly optimize these three control dimensions, we propose~\ourSystem{}, a carbon-aware adaptive control framework for edge~AI.
Proactive decision-making requires anticipating variable grid conditions, yet edge devices typically lack sufficient history for forecaster training.
\ourSystem{} addresses this limitation by leveraging edge-friendly~Time-Series Foundation Models~(TSFMs)~\cite{goswami_moment_2024, ansari2024chronos, arxiv_carbonx, quan2025benchmarking},~\eg~Sundial~\cite{liu2025sundial}, for periodic zero-shot multi-step forecasting.
These forecasts are then passed to our receding-horizon dynamic programming~(DP) solver for joint optimization. 
However, truncating the optimization horizon introduces a key failure mode,~\ie~the~\emph{free-discharge loophole}: because carbon emissions are incurred only during battery charging, the solver may discharge the battery myopically and defer the resulting recharge to a later high-carbon period. 
To mitigate this issue, \ourSystem{} introduces a deferred cost attribution mechanism that internalizes the anticipated carbon cost of future recharge into the current discharge decision.

We implement~\ourSystem{} on an NVIDIA Jetson Orin Nano with ten-variant object detection and classification pipelines and up to~60~hardware configurations, driven by three years of carbon-intensity traces across six regions. 
Results show that \ourSystem{} reduces carbon emissions by up to~65.6\% relative to a carbon-unaware baseline while maintaining near-maximum inference accuracy.

\section{Background and Related Work}
\label{sec:related_work}

\textbf{Edge~AI.}
Recent advances in~AI accelerators, including~NVIDIA~Jetson platforms, NPUs, and Google~TPUs~\cite{nvidia_jetson,TPUPatent,Tan2022:NPUs}, have made it increasingly practical to deploy~AI models on edge devices~\cite{Dong2025:EdgeAI}.
These systems often process data from local sensors or provide inference services to nearby resource-constrained clients. 
Compared with server-class infrastructure, however, edge accelerators operate under much tighter power budgets. 
For example, the~NVIDIA~Jetson Orin~Nano has a maximum power draw of only~15~W~\cite{nvidia_jetson}. 
This modest power demand, together with declining battery costs~\cite{battery_drop}, makes it practical to equip edge devices with embedded batteries for backup resilience and off-grid operation~\cite{hasan2023battery,Mousavi2017:Accuracy-resource}.

To clarify the scope of~\ourSystem{}, we distinguish three edge battery paradigms.
\emph{1)~Grid-connected and battery-equipped devices.} 
This is a natural target for~\ourSystem{}, including commercial products such as security cameras, \eg Nest~Cam with Floodlight~\cite{nest_cam_floodlight}, and industrial gateways, \eg TEKTELIC~KONA~Micro~\cite{tektelic_kona_micro}. 
\emph{2)~Grid-only devices.}
Systems such as smart displays could also support~\ourSystem{} after battery integration, but at the cost of additional embodied carbon~\cite{peiseler2024carbon}.
\emph{3)~Battery-only devices.} 
Mobile devices, such as smartphones, face user recharge uncertainty, requiring behavior-aware charging.

\textbf{Energy- and Carbon-Aware Systems.}
Prior work on edge~AI has largely focused on improving energy efficiency. 
These approaches navigate accuracy and energy trade-offs through dynamic model inference, hardware adaptation, and QoS-aware scheduling~\cite{bullo2024energy,zhang2025e4,jeon2023harvnet,han2025dvfs,dai2024energy,siar2025energy,ma2025greening,Liang2023:Delen,chen2026greener}.
While effective at reducing energy consumption, they remain carbon-agnostic. 
Meanwhile, carbon-aware computing has gained traction in cloud datacenters and broader cyber-physical systems,~\eg~EV charging and smart buildings. 
These approaches reduce emissions by exploiting \emph{spatiotemporal workload flexibility}, either by shifting execution across regions with cleaner grids or by deferring tasks until carbon intensity decreases~\cite{radovanovic2022carbon,wiesner2021:letswait,maji2023multicloud,jiang2024ecolife,Wu2025:CarbonEdge,lechowicz2025online,su2023learning,faruqui2011smartprices,jones2022tou,cheng2022carbonawareev,dixon2020evcharging,li2023mef_tou,Bovornkeeratiroj2025:GreenThrift, hanafy2023carbonscaler}. 
Directly applying these strategies to edge~AI is challenging because always-on edge inference must satisfy strict real-time and data-locality constraints, leaving little room for delay or offloading. 
Instead of relying on external spatiotemporal flexibility, \ourSystem{} exploits three sources of internal elasticity in edge systems to reduce carbon without violating~QoS constraints.

\section{Design of \ourSystem{}}
\label{sec:design}

\ourSystem{} targets a grid-connected edge device,~\eg~a smart security camera, equipped with a battery.
The goal is to maximize QoS~(task accuracy and latency) while minimizing carbon emissions and electricity costs.
At each interval~\( t \), a controller observes the system state~\( s_t = \left(b_t, g_t, p_t\right) \), which includes the battery energy level~\( b_t \), grid carbon intensity~\( g_t \), and electricity price~\( p_t \).
Based on this information, the controller determines an action~\( a_t = \left(o_t, c_t, \delta_t\right) \), where~\( o_t = \left(n_t, h_t\right) \) represents the operating mode~(task model variant and hardware configuration),~\( c_t \) is the battery charging decision, and~\( \delta_t \in \left\{\textsc{Battery}, \textsc{Grid}\right\} \) is the power source for inference.

Achieving favorable performance requires balancing actions to meet application and environmental goals. 
\S\ref{sec:design:objective} defines the control objective to maximize long-term~QoS while minimizing carbon emissions and electricity costs. 
\S\ref{sec:holisticsystem} formalizes the state and action spaces with a battery-buffered model, and~\S\ref{sec:design:algorithm} introduces the predictive controller to solve the optimization problem online.

\subsection{Problem Formulation}
\label{sec:design:objective}

We optimize long-term~\(T\)-step control by balancing mode performance utility~\(U^{\text{perf}}_t\), carbon emissions~\(C^{\text{carb}}_t\), and electricity cost~\(C^{\text{cost}}_t\). 
At each control step~\(t\), the policy~\(\pi\) selects an action~\(a_t = \left(o_t, c_t, \delta_t\right)\), where~\(o_t = \left(n_t, h_t\right)\) is the operating mode~(model variant and hardware configuration),~\(c_t \in \left\{0, 1\right\}\) is the binary charging decision, and~\(\delta_t \in \left\{\textsc{Battery}, \textsc{Grid}\right\}\) is the power source for inference:
\begin{equation}
\begin{aligned}
\max_{a_t \in \pi}\quad & \mathbb{E}\left[\sum_{t=0}^{T-1} w_{\text{perf}} U^{\text{perf}}_t\left(a_t\right) - w_{\text{carb}} C^{\text{carb}}_t\left(a_t\right) - w_{\text{cost}} C^{\text{cost}}_t\left(a_t\right) \right] \\
\text{s.t.}\quad & \alpha\left(n_t\right) \ge u_{\text{acc}}, \quad \ell\left(n_t,h_t\right) \le u_{\text{lat}},
\end{aligned}
\label{eq:objective}
\end{equation}
where~\(w_{\text{perf}}, w_{\text{carb}}, w_{\text{cost}} \ge 0\) are weights. 
The chosen operating mode’s accuracy~\(\alpha\left(n_t\right)\) and latency~\(\ell\left(n_t, h_t\right)\) should meet the strict thresholds~\(u_{\text{acc}}\) and~\(u_{\text{lat}}\), respectively. 
However, simply meeting these constraints is not sufficient; the primary goal is to maximize them while minimizing operational emissions and costs.

\textbf{Performance Utility~\( U^{\text{perf}}_t \).}
Each selected operating mode~\( o_t = \left(n_t, h_t\right) \) induces an accuracy-latency pair,~\ie~\( \alpha\left(n_t\right) \) and \( \ell\left(n_t, h_t\right) \). 
For QoS-feasible modes, the utility is defined as a margin-based reward relative to the accuracy and latency requirements:
\begin{equation}
U^{\text{perf}}_t = \left[\alpha\left(n_t\right) - u_{\text{acc}}\right]^+ + \lambda_{\ell}\left[\frac{1}{\ell\left(n_t,h_t\right)} - \frac{1}{u_{\text{lat}}}\right]^+,
\label{eq:uperf}
\end{equation}
where~\( \left[x\right]^+ = \max\left\{x, 0\right\} \) and~\( \lambda_{\ell}\) controls the relative importance of latency. 
This formulation rewards modes that exceed the minimum accuracy requirement and improve latency beyond the threshold.

\textbf{Carbon Emissions~\( C^{\text{carb}}_t \) and Electricity Cost~\( C^{\text{cost}}_t \).}
Let~\( g_t \) and~\( p_t \) denote the grid carbon intensity~(gCO\(_2\)/kWh) and electricity price~(\$/kWh), respectively. 
Both carbon emissions and electricity cost are attributed at the point of grid draw:
\begin{equation}
\begin{bmatrix}
C^{\text{carb}}_t \\ 
C^{\text{cost}}_t 
\end{bmatrix} 
= 
\begin{bmatrix}
g_t \\ 
p_t 
\end{bmatrix} 
\left( c_t \cdot P^{\text{chg}} \cdot \Delta t + \mathbb{I}\left[\delta_t=\textsc{Grid}\right]\, E^{\text{infer, raw}}_t\left(o_t\right) \right),
\end{equation}
where~\( c_t \cdot P^{\text{chg}} \cdot \Delta t \) is the total energy drawn from the grid for battery charging during the interval~$\Delta t$, and~\( \mathbb{I}\left[\delta_t=\textsc{Grid}\right]\, E^{\text{infer, raw}}_t\left(o_t\right) \) is the energy drawn from the grid for the inference workload~(\S\ref{sec:design:energymodel}).

\subsection{Battery-Buffered System Modeling}
\label{sec:holisticsystem}
We define the system modeling space of the edge device in three aspects: available inference models, feasible hardware configurations for running them, and the power source~(battery or grid).

\subsubsection{Inference Model~\(\left(\mathcal{N}\right)\)}
\label{sec:design:modelspace}

We consider an edge device running a compound AI service,~\ie~a multi-stage pipeline such as trigger, detection, tracking, and refinement.
At each stage, the device can choose from multiple model variants, and the selected models define a pipeline configuration.
Let~$\mathcal{S} = \left\{1, \dots, S\right\}$ index the stages, and~$\mathcal{M}_s$ denote the set of model variants available at stage~$s$. 
A pipeline configuration is an~$S$-tuple~$n = \left(m_1, \dots, m_S\right)$, where~$m_s \in \mathcal{M}_s$. 
The inference model space~$\mathcal{N}$ is the set of all feasible configurations:
\begin{equation}
\mathcal{N} \subseteq \mathcal{M}_1 \times \cdots \times \mathcal{M}_S,
\end{equation}
where each~$n \in \mathcal{N}$ represents a combination of models, with an associated benchmarked accuracy score~$\alpha\left(n\right)$. 
It is typically evaluated on a representative dataset, such as mean average precision~(mAP50-95) on the COCO benchmark for object detection~\cite{lin2014microsoft}.

\subsubsection{Hardware Configuration \(\left(\mathcal{H}\right)\)}
\label{sec:design:hardwarespace}

Let~\(\mathcal{H}\) denote a set of feasible hardware operating points exposed by the edge platform. 
We represent a hardware state~\(h \in \mathcal{H}\) as:
\begin{equation}
    h = \left\langle \tau_{\text{unit}}, f_{\text{clk}}, c_{\text{cores}}, p_{\text{cap}} \right\rangle ,
\end{equation}
where~\(\tau_{\text{unit}} \in \left\{\text{CPU}, \text{GPU}, \text{NPU}, \ldots\right\}\) selects the execution unit,~\(f_{\text{clk}}\)~sets the dynamic voltage and frequency scaling~(DVFS) level for the unit,~\(c_{\text{cores}}\) specifies the number of active cores, and~\(p_{\text{cap}}\) is the platform power mode, which fundamentally limits the allowable clock settings and active resources~\cite{nvidia_jetson}.
The joint selection of these parameters determines the runtime latency and energy consumption for a given pipeline configuration, defining the hardware elasticity available to the controller~\cite{nvidia_jetson}.

\textbf{Unified Operational Mode Space \(\left(\mathcal{O}\right)\).}
An \emph{operational mode} is defined as the joint selection of an inference pipeline configuration~\(n \in \mathcal{N}\) and a hardware state~\(h \in \mathcal{H}\). 
The mode set is
\begin{equation}
    \mathcal{O} = \mathcal{N} \times \mathcal{H}.
\end{equation}
At each step~\(t\), the controller selects~\(o_t = \left(n_t, h_t\right) \in \mathcal{O}\), which induces the profiled attribute tuple:
\begin{equation}
    \text{Attr}\left(o_t\right) = \left\langle \alpha\left(n_t\right), \ell\left(n_t, h_t\right), e\left(n_t, h_t\right) \right\rangle ,
\end{equation}
where~\(\alpha\left(n_t\right)\) is the benchmarked accuracy score,~\(\ell\left(n_t, h_t\right)\) is the inference latency under the selected hardware state~\(h_t\), and~\(e\left(n_t, h_t\right)\) is the corresponding inference energy cost obtained from profiling.

\subsubsection{Battery Modeling}
\label{sec:design:energymodel}

Real-world battery dynamics involve complex non-linearities, such as temperature dependencies and transient voltage drops~\cite{rahimi2013battery}. 
While these electrochemical phenomena are crucial for Battery Management Systems, modeling them is computationally prohibitive for edge control. 
Thus, we abstract these transients using a constant nominal voltage~\(V^{\text{nom}}\) and charging efficiency~\(\eta_{\text{chg}}\), while retaining two key constraints: rate-dependent capacity loss during high-power inference and capacity degradation limits via a state of charge~(SoC) bounding window.

Let~\(B^{\text{cap}}\) represent the maximum battery capacity, and~\(b_t\) denote the available battery energy at step~\(t\).
Battery energy evolves as:
\begin{equation}
b_{t+1} = \min\left\{B^{\text{cap}},\; \left[b_t + E^{\text{in, eff}}_t - E^{\text{out, eff}}_t \left(o_t\right)\right]_+\right\},
\label{eq:batt_dyn}
\end{equation}
where~\(E^{\text{in, eff}}_t\) is the effective charging energy stored from the grid, and~\(\mathbb{I}\left[\delta_t=\textsc{Battery}\right]\,E^{\text{out, eff}}_t\left(o_t\right)\) is the effective energy discharged from the battery when it is selected as the power source. 
The notation~\(\left[ x \right]_+ = \max\left\{x, 0\right\}\) ensures that the battery state cannot drop below zero, while the~\(\min\) function prevents the stored energy from exceeding capacity~\(B^{\text{cap}}\).
To promote long-term battery health, the controller is constrained to operate within a 20\%--80\%~SoC window, enforced by the energy bounds~\(b_t \in \left[0.2B^{\text{cap}},\, 0.8B^{\text{cap}}\right]\).

\textbf{Charging Energy~\(E^{\text{in, eff}}_t\).}
For simplicity, we do not directly control the amount of energy charged, but instead assume the battery is charged throughout the entire control interval~$\Delta t$, based on a binary decision~$c_t \in \left\{0, 1\right\}$.
If \(c_t = 1\), the charger draws grid power~\(P^{\text{chg}}\) for the duration of the control interval~\(\Delta t\). 
Accounting for the charging efficiency~\(\eta_{\text{chg}}\), the grid-to-battery inflow is:
\begin{equation}
E^{\text{in, eff}}_t = \min\left\{c_t \cdot P^{\text{chg}} \cdot \Delta t \cdot \eta_{\text{chg}},\, B^{\text{cap}} - b_t \right\}.
\label{eq:ein_eff}
\end{equation}
If the total energy directed to the battery during the interval exceeds the remaining headroom \(\left(B^{\text{cap}} - b_t\right)\), it is capped.

\textbf{Discharging Energy~\(E^{\text{out, eff}}_t\).}
We first compute the inference energy consumption~\(E^{\text{infer, raw}}_t\) at each control interval given the operating mode~\(o_t\), then determine the corresponding consumed battery energy~\(E^{\text{out, eff}}_t\) if~\(\delta_t = \textsc{Battery}\), considering the non-linear battery discharge behavior modeled by Peukert's law~\cite{cugnet2010peukert}.

\textsc{(i)~Inference Energy~\(E^{\text{infer, raw}}_t\).} 
Given the selected mode~\(o_t = \left(n_t, h_t\right)\) and the number of executed inferences~\(N_t^{\text{infer}}\) during the control interval, the raw inference energy is:
\begin{equation}
E^{\text{infer, raw}}_t\left(o_t\right) = N_t^{\text{infer}} \cdot e\left(n_t, h_t\right),
\label{eq:eraw}
\end{equation}
where~\(e\left(n_t, h_t\right)\) is the profiled energy per inference.

\textsc{(ii)~Peukert's Law-Guided Battery Discharge.}
If~\(\delta_t = \textsc{Battery}\), a discharge penalty is applied to account for rate effects. 
Let~\(V^{\text{nom}}\) denote the nominal voltage. 
The discharge current~\(i_t\) and the rate-dependent penalty factor~\(\eta\left(i_t\right)\) are computed as:
\begin{equation}
i_t = \frac{E^{\text{infer, raw}}_t\left(o_t\right)}{1000 \cdot \Delta t \cdot V^{\text{nom}}}, \qquad \eta\left(i_t\right) = \left(\frac{i_t}{i^{\text{ref}}}\right)^{k-1},
\end{equation}
where~\(k > 1\) is the Peukert exponent that captures rate-dependent capacity loss, and~\(i^{\text{ref}}\) is the reference current of the battery.
The effective discharge energy is then given by:
\begin{equation}
E^{\text{out, eff}}_t\left(o_t\right) = \eta\left(i_t\right) \cdot E^{\text{infer, raw}}_t\left(o_t\right).
\end{equation}
When the inference workload is served from the grid, no battery penalty is applied, and the grid supplies~\(E^{\text{infer, raw}}_t\left(o_t\right)\) directly.

\begin{algorithm}[t]
\caption{\ourSystem{} Controller Policy}
\label{alg:tsfm-mpc}
\begin{lstlisting}[language=Python]
# s_t = (b_t, g_t, p_t)        : state
# a_t = (mode, charge, source) : action
# H:      MPC look-ahead horizon
# K:      TSFM reforecast interval (K = H)
# T_cold: cold-start duration
\end{lstlisting}
\begin{lstlisting}[language=Python]
for t in range(T):
    g_ci.append(g_t); p_ci.append(p_t)
    if t < T_cold:
        a_t = carbon_threshold(s_t); continue
    if no_forecast or (t - t_last) >= K:
        g_hat, g_std = tsfm_forecast(g_ci, 2*H)
        p_hat, p_std = tsfm_forecast(p_ci, 2*H)
        t_last = t
\end{lstlisting}
\begin{lstlisting}[language=Python]
    i   = t - t_last
    g_f = g_hat[i : i + H]
    p_f = p_hat[i : i + H]
    rho = confidence(g_std[i:i+H], p_std[i:i+H])
    Pi  = dp_solve(g_f, p_f, rho)
    a_t = lookup(Pi, k=0, b_t)
    s_t = env.step(a_t)
\end{lstlisting}
\end{algorithm}

\subsection{TSFM-Driven Predictive Control}
\label{sec:design:algorithm}

Our controller relies on carbon intensity and cost forecasts to enable anticipatory decision-making and online adaptation under evolving grid conditions.
To achieve this capability without on-device training, we leverage~TSFMs, \ie large-scale models pre-trained on diverse time-series corpora~\cite{goswami_moment_2024, ansari2024chronos, wang2025spectral}. 
This design is well-suited to edge settings, where sufficient local data for training a forecasting model is often unavailable. 
By learning transferable temporal representations,~TSFMs support robust zero-shot forecasting across heterogeneous temporal patterns. 
Building on this capability, we formulate a~TSFM-based model predictive control framework.

\textbf{TSFM Forecasting and Model Predictive Control~(MPC).}
Algorithm~\ref{alg:tsfm-mpc} summarizes the control loop.
At each step~$t$, \ourSystem{} records the observed carbon intensity~$g_t$ and electricity price~$p_t$.
During cold start, the controller follows a carbon-threshold heuristic~(\S\ref{sec:eval:baseline}) to collect sufficient historical context. 
Once this phase ends, the~TSFM is queried every~$K = H$ steps to produce a probabilistic~$2H$-step forecast for carbon~$\left(\hat{g}, \sigma_g\right)$ and price~$\left(\hat{p}, \sigma_p\right)$.
This maintains a valid~$H$-step look-ahead window throughout each reforecast interval through the sliding index~$i = t - t_{\text{last}}$.

At each step, a confidence score~$\rho$ is computed from temporal decay and forecast variance~$\sigma$. 
Since forecast reliability decreases with horizon,~$\rho$ acts as a calibration factor in the objective, down-weighting penalties associated with uncertain long-horizon predictions and placing greater emphasis on reliable near-term information.
The resulting forecast slices and confidence values are then passed to~\textsc{DPSolve}, our~DP routine over a discretized battery state space, to compute a local control policy~$\Pi_t$. 
Since~$\Pi_t$ is derived from imperfect forecast information, \ourSystem{} executes only the action at step~$t$ and re-optimizes at the next step, thereby enabling responsive online adaptation to realized grid conditions.

\textbf{DP Solver with Deferred Attribution.}
While the receding-horizon framework provides necessary real-time adaptability, it limits the optimization scope. 
\textsc{DPSolve} performs backward induction over a discretized battery state space within a rolling finite horizon.
Ideally, the~DP would optimize over the full episode with complete future knowledge. 
However, this is infeasible due to computational overhead and the degradation of long-term forecasts, which both scale with episode length.
Thus, the~DP operates over a truncated receding horizon, introducing a critical vulnerability,~\ie~the \emph{free-discharge loophole}. 
Because carbon emissions and costs are exclusively incurred at the moment of grid charging, drawing power from the battery incurs no immediate penalty.
With a shorter DP horizon, any battery depletion that pushes the inevitable recharge event beyond the visible window appears entirely costless to the solver. 
The DP myopically drains the battery to support high-accuracy models, forcing a later recharge, potentially during peak carbon intensity, undermining carbon-aware buffering.

To close this \emph{free-discharge loophole}, \ourSystem{} introduces a novel \emph{deferred cost attribution} mechanism. 
Instead of treating battery discharge as free, the controller anticipates future recharge events and applies a deferred penalty based on forecasted grid conditions beyond the current horizon. 
This penalty is estimated using the~$q$-th percentile of the remaining forecast, reflecting the assumption that the controller will recharge during the lowest-carbon and lowest-cost window available within the remaining forecast.
By internalizing the full lifecycle carbon footprint, this mechanism ensures the battery is drained only when current grid conditions are worse in carbon intensity and cost than future recharge opportunities.

\section{Implementation}
\label{sec:implementation}

This section focuses on the security camera scenario with its default settings listed. 
Varying settings and scenarios are covered in \S\ref{sec:deepdive}.

\begin{figure*}[t]
  \centering
  \begin{subfigure}[b]{0.24\textwidth}
    \centering
    \includegraphics[width=\textwidth]{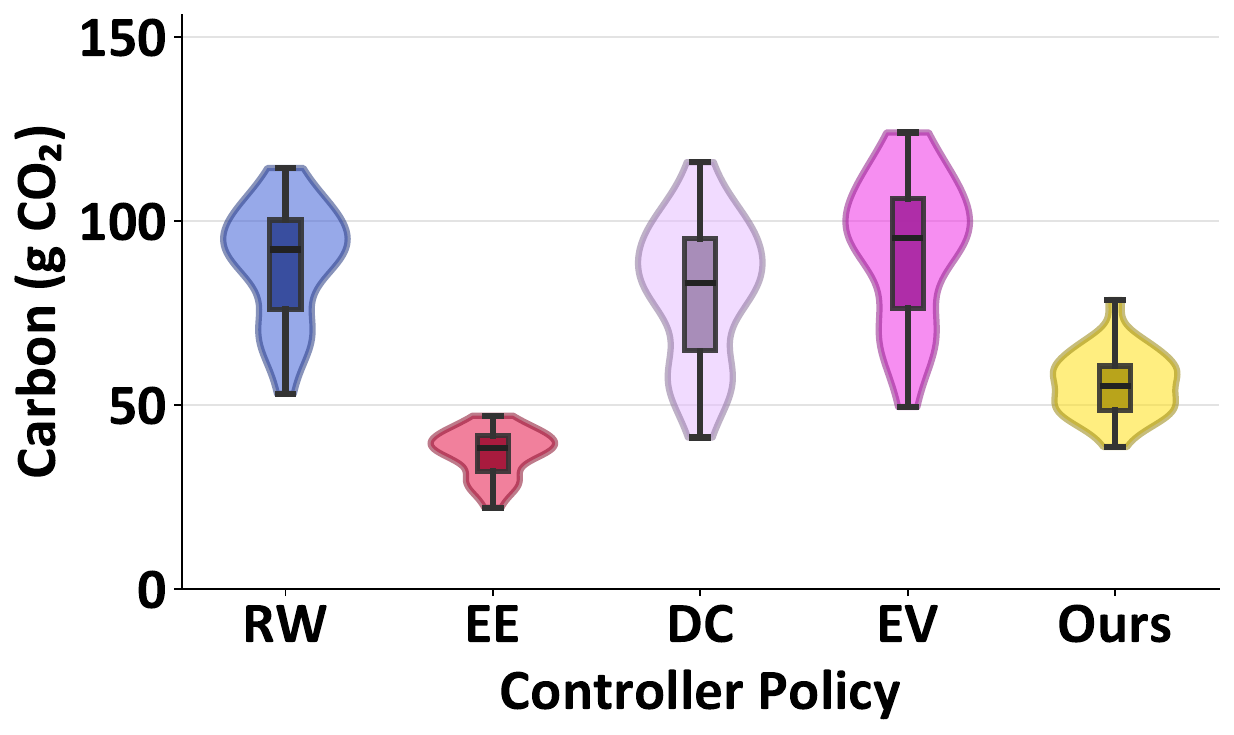}
    \caption{Carbon emissions}
    \label{fig_overall_a}
  \end{subfigure}
  \hfill
  \begin{subfigure}[b]{0.24\textwidth}
    \centering
    \includegraphics[width=\textwidth]{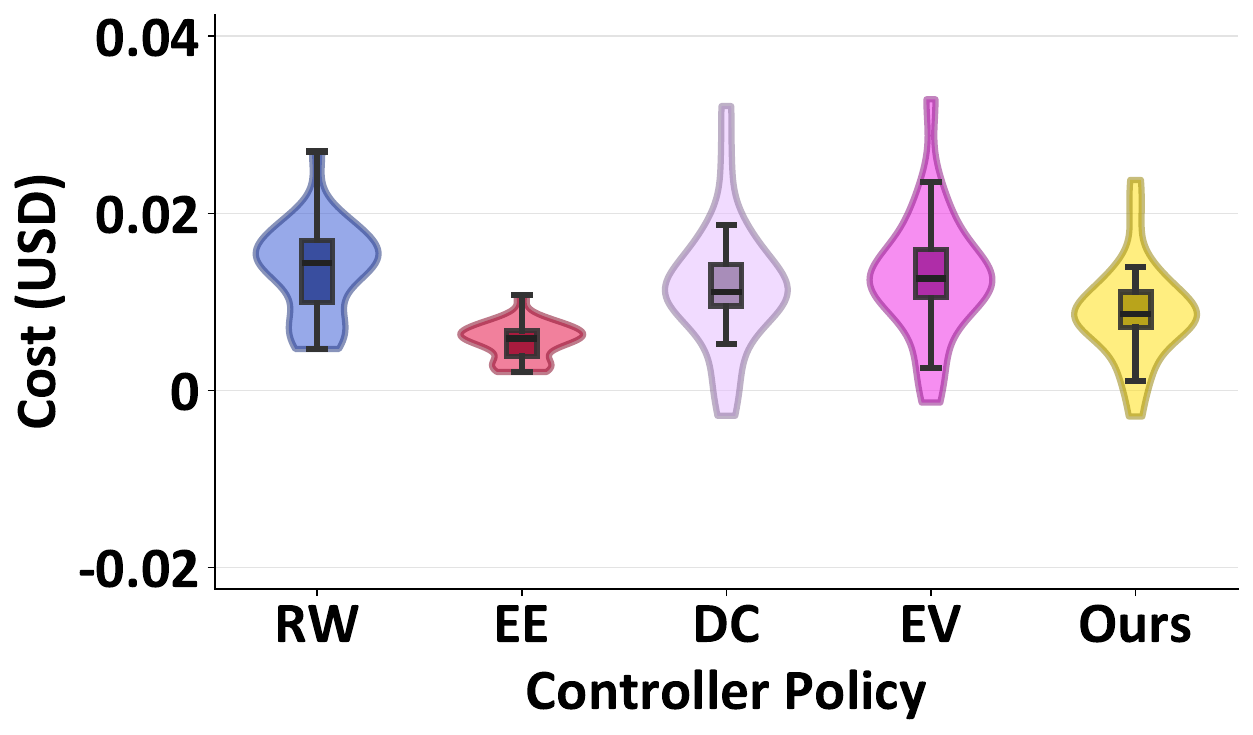}
    \caption{Electricity cost}
    \label{fig_overall_b}
  \end{subfigure}
  \hfill
  \begin{subfigure}[b]{0.24\textwidth}
    \centering
    \includegraphics[width=\textwidth]{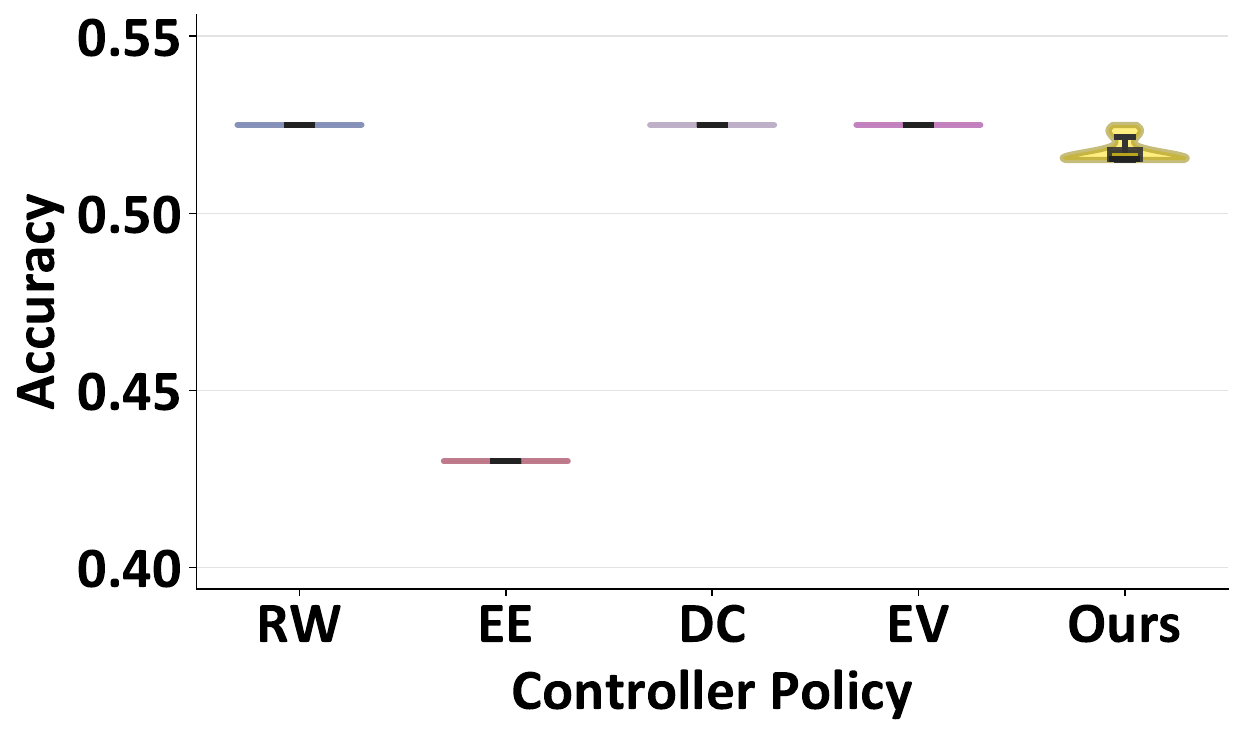}
    \caption{Inference accuracy~(mAP50--95)}
    \label{fig_overall_c}
  \end{subfigure}
  \hfill
  \begin{subfigure}[b]{0.24\textwidth}
    \centering
    \includegraphics[width=\textwidth]{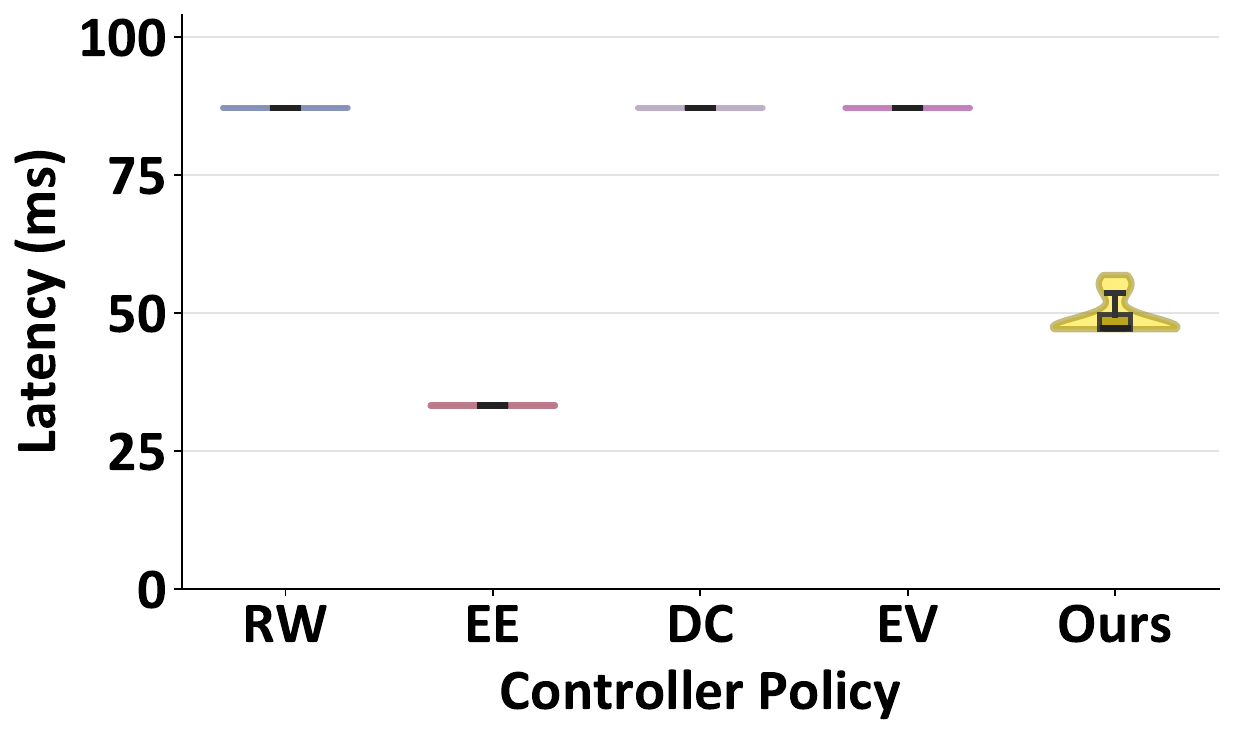}
    \caption{Inference latency}
    \label{fig_overall_d}
  \end{subfigure}
\caption{Comparison of~\ourSystem{}~(Ours) and four baselines across four key metrics in the CAISO region over the test period.}
\Description[]{}
  \label{fig_overall}
\end{figure*}

\textbf{Edge System.}  
We target a smart security camera scenario running on an NVIDIA Jetson Orin Nano~8GB. 
The camera operates at a rate of 1.0 inference per second. 
The inference pipeline performs object detection using ten~\texttt{YOLO} model variants, ranging from lightweight to high-accuracy: \texttt{YOLOv5nu}, \texttt{YOLOv8n}, \texttt{YOLO11n}, \texttt{YOLO12n}, \texttt{YOLOv5su}, \texttt{YOLOv10s}, \texttt{YOLOv8m}, \texttt{YOLO11m}, \texttt{YOLO12m}, and \texttt{YOLOv9e}.
The mAP50-95 scores for these models range from~0.343 to~0.556 on the \texttt{COCO val2017} benchmark dataset~\cite{lin2014microsoft}.

\textbf{Hardware.}
We profile~60~hardware configurations formed by combining five~GPU clock levels~(306--625\,MHz), two~CPU frequencies~(883 and 1510\,MHz), and six active-core counts~(1--6).
For each of the ten~\texttt{YOLO} models under each hardware configuration, we measure per-inference latency and average power consumption on 640$\times$640 input images, averaged over~100~requests.
The measured latency ranges from~30.9\,ms to~235.6\,ms, while average power ranges from~4.1\,W to~10.0\,W.
This profiling defines a joint mode space~$\mathcal{O} = \mathcal{N} \times \mathcal{H}$ with~600~distinct model--hardware pairs. 
We further impose strict~QoS constraints, requiring a minimum accuracy of~$u_{\text{acc}} = 0.40$~mAP50--95 and a maximum latency of~$u_{\text{lat}} = 100$\,ms.

\textbf{Battery.}
We model a compact consumer-grade battery~\cite{iphone_battery} with capacity~$B^{\text{cap}} = 18{,}000$\,mWh. 
The charger provides~$P^{\text{chg}} = 20$\,W at efficiency~$\eta_{\text{chg}} = 0.90$. 
We use a Peukert discharge model with exponent~$k = 1.05$ to capture rate-dependent capacity loss.

\textbf{\ourSystem{}.}
The TSFM uses \texttt{sundial-base-128m}~\cite{liu2025sundial} with $N_s = 20$ probabilistic samples, reforecasted every $K = 96$ slots using up to 1{,}344 slots~(14 days) of historical data as context. 
The~DP look-ahead is $H = 96$ slots with $L = 100$ battery levels and a discount factor of $\gamma = 0.998$. 
The cold-start phase runs for $T_{\text{cold}} = 96$ slots, during which the data centers carbon-aware policy~(\S\ref{sec:eval:baseline}) is used.

\section{Evaluation}
\label{sec:evaluation}

We evaluate the overall performance of \ourSystem{} in~\S\ref{sec:eval:overall}, followed by the generalization study in~\S\ref{sec:deepdive}, the ablation study in~\S\ref{sec:ablation}, and finally, the overhead analysis in \S\ref{sec:overhead}.

\subsection{Experimental Setting}

\subsubsection{Carbon Intensity and Electricity Price Trace-Driven Dataset.}
We collect three years of grid carbon intensity data from Electricity Maps and electricity price data via the Python \texttt{gridstatus} package at 15-minute resolution across six regions: CAISO, PJM, ERCOT, ISO-NE, BPAT, and NYISO.
Each episode spans $T = 2{,}880$ control slots, corresponding to 30 days at $\Delta t = 0.25$\,h per slot.
The dataset is split temporally into 30\% validation and 70\% testing.

\subsubsection{Four Performance Criteria}
We evaluate each controller across the full test period using four episode-level metrics:
{1)~Carbon Emissions~(g\,CO$_2$)} measures total consumed CO$_2$; 
{2)~Electricity Cost~(USD)} measures total grid electricity expenditure per episode; 
{3)~Inference Accuracy} is the episode-average task accuracy; 
and~{4)~Inference Latency~(ms)} is the episode-average per-inference latency. 
A good controller minimizes carbon and cost, while maximizing accuracy and minimizing latency.

\subsubsection{Baselines}
\label{sec:eval:baseline}

We compare \ourSystem against four baselines, representing strategies from performance-oriented to energy-aware, EV charging, and data center-inspired carbon-aware approaches.

\begin{itemize}[label=\textbullet, leftmargin=1.5em]

\item \textbf{Real-World Deployment (RW)~\cite{alqahtani2024benchmarking}.}  
This baseline represents a \emph{real-world, energy-unaware deployment} that prioritizes \emph{highest performance}. 
It always uses \textsc{Grid} and selects the highest-accuracy feasible mode.
This baseline exemplifies a typical carbon-unaware, always-on edge deployment.

\item \textbf{Energy-Efficiency~(EE)~\cite{zhang2025e4}.}
It~\emph{minimizes energy consumption} by selecting the lowest-energy mode that remains~\emph{QoS-feasible}, while always using the \textsc{Grid}.
It shows that purely minimizing energy severely degrades other system aspects, \eg accuracy.

\item \textbf{Data Centers Carbon-Aware Policy (DC)~\cite{wiesner2021:letswait}.}  
This baseline mimics widely adopted workload shifting policies in data centers, using a \emph{rolling history of carbon intensity} to make decisions. 
It charges the battery from the \textsc{Grid} when carbon intensity is low and powers the inference workload from the \textsc{Battery} when carbon intensity is high. 
It selects the best-performance feasible mode under the chosen source, without workload adaptation.

\item \textbf{EV Charging Policy (EV)~\cite{dixon2020evcharging}.}
It remains idle until the~SoC falls below a preset floor, then waits for a low-carbon-intensity window to begin charging, and continues charging until the~SoC reaches a target level. It always selects the highest-performance feasible mode, without workload adaptation.

\end{itemize}

\subsection{Overall Performance}
\label{sec:eval:overall}

Figure~\ref{fig_overall} compares~\ourSystem{} with the baselines across four metrics over the test period in the~CAISO region. 
Overall, \ourSystem{} achieves the second-lowest carbon emissions and electricity cost while remaining close to the best inference accuracy and latency. 
Note that negative electricity costs can occur during periods of renewable overgeneration, when market prices drop below zero.

\textbf{RW} incurs carbon emissions of~89.1\,gCO$_2$ and the highest cost of~\$0.014 among all policies. Relative to~RW, \ourSystem{} reduces carbon emissions by~37.3\% and cost by~36.0\%, with only a negligible accuracy drop from~0.525 to~0.518\,mAP and a latency reduction from~87.2\,ms to~49.4\,ms. 
These results show that forecast-driven energy management avoids the carbon and cost penalties of always-on grid operation without sacrificing service quality.

\textbf{EE} achieves the lowest carbon emissions, but also the lowest accuracy.
Compared with~EE, \ourSystem{} incurs~18.9\,gCO$_2$ more carbon while improving accuracy by~0.088\,mAP. 
This highlights that energy-efficient policies can substantially sacrifice inference quality for carbon reduction, whereas \ourSystem{} achieves a more balanced trade-off across all four metrics.

\textbf{DC} maintains full accuracy at~0.525\,mAP but incurs~81.3\,gCO$_2$, corresponding to~45.6\% higher carbon emissions than~\ourSystem{}. 
The~DC policy exhibits repetitive oscillations between the upper and lower SoC bounds because it reacts only to instantaneous carbon-intensity thresholds. 
As a result, it cannot exploit future low-carbon windows, and its fixed mode selection misses the additional savings captured by \ourSystem{}'s joint forecast-driven optimization.

\textbf{EV} incurs the highest carbon emissions at~93.2\,gCO$_2$. 
Compared with~EV, \ourSystem{} reduces carbon emissions by~40.0\%, cost by~31.7\%, and latency by~43.3\%, from~87.2\,ms to~49.4\,ms, through coordinated mode selection. 
This indicates that EV-style policies are poorly suited to edge inference workloads.

\subsection{Generalization Study}
\label{sec:deepdive}

\subsubsection{Second Scenario \textendash{} Image Classification}
\label{sec:eval:image}

To evaluate generalization beyond object detection, we apply~\ourSystem{} to an image classification scenario on the same NVIDIA Jetson Orin Nano~8GB platform. 
The pipeline includes ten~\texttt{torchvision} models, ranging from~ResNet-18 to~ViT variants. 
The controller selects from~30~hardware configurations over CPU core counts and CPU/GPU frequencies, yielding~300~profiled model--hardware pairs. 
Measured latencies range from~9.0\,ms to~229.1\,ms. 
The~QoS constraints require a minimum top-1 accuracy of~0.75 and a maximum latency of~100\,ms. 
All other settings follow the primary scenario.

\begin{figure}[t]
  \centering
  \begin{subfigure}[b]{0.49\columnwidth}
    \centering
    \includegraphics[width=\linewidth]{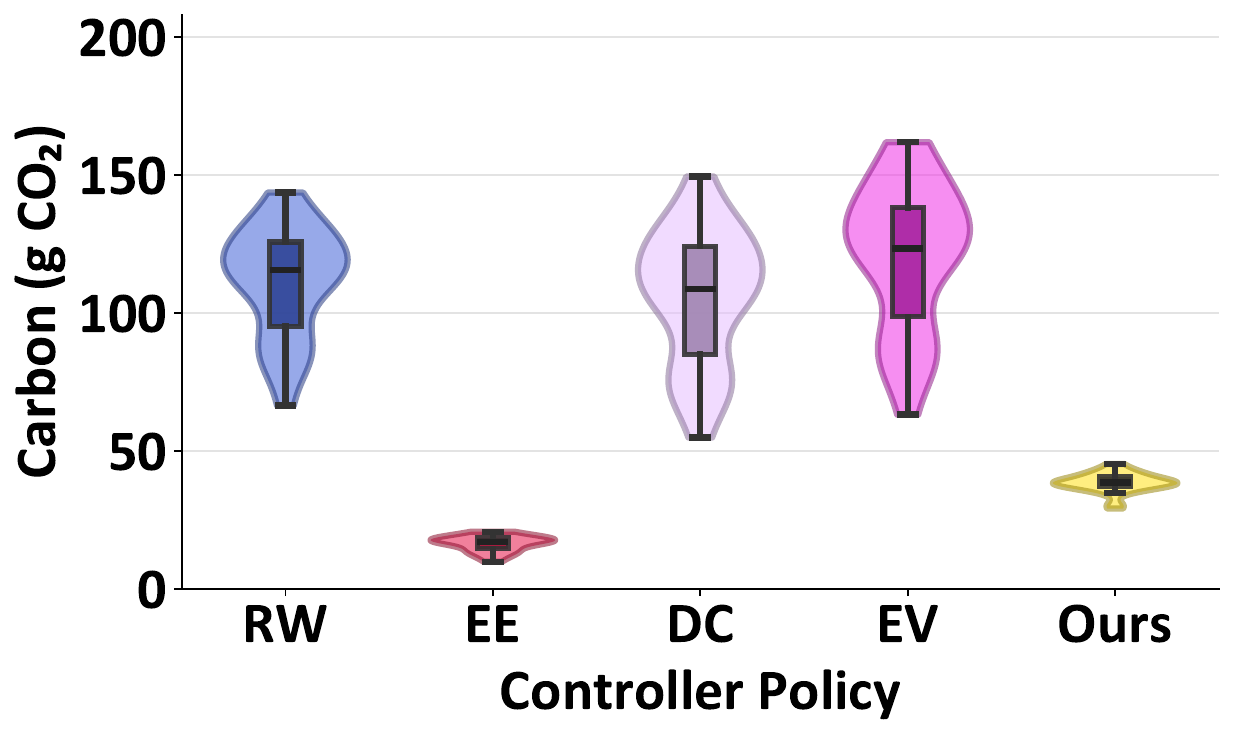}
    \caption{Carbon emissions}
    \label{fig_carbon_classification}
  \end{subfigure}
  \hfill
  \begin{subfigure}[b]{0.49\columnwidth}
    \centering
    \includegraphics[width=\linewidth]{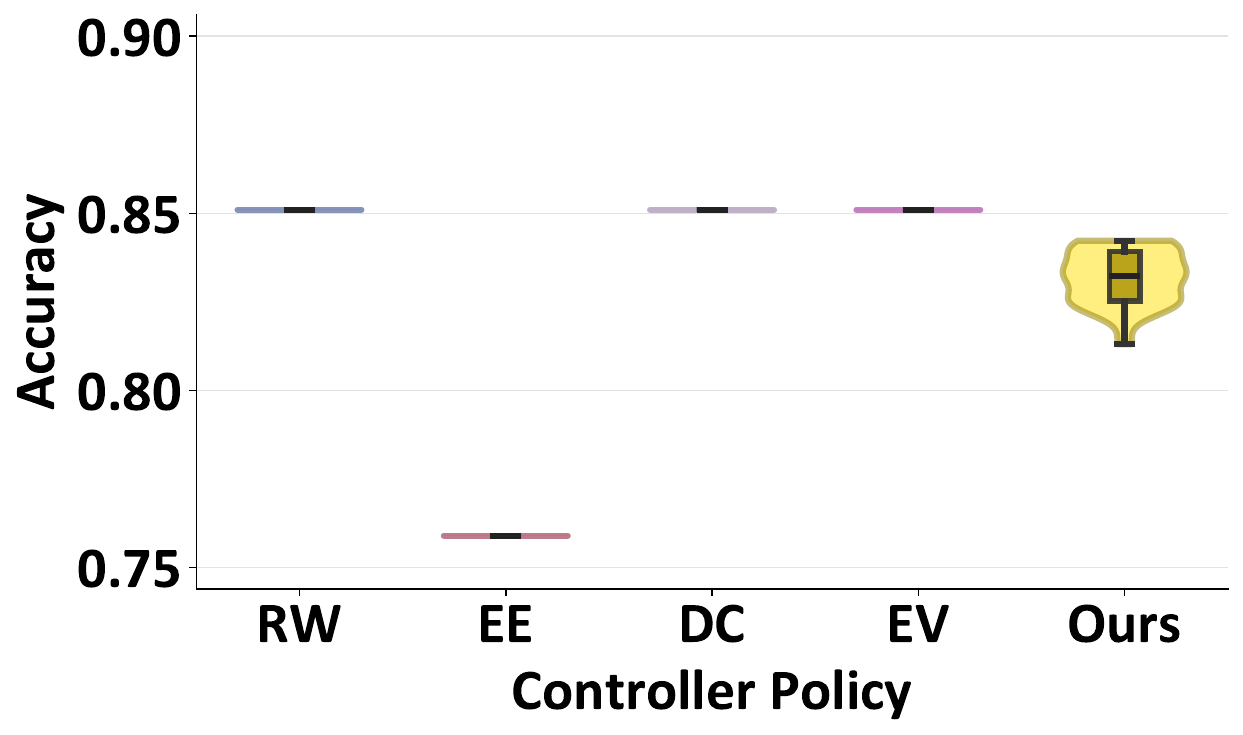}
    \caption{Inference accuracy~(Acc@1)}
    \label{fig_accuracy_classification}
  \end{subfigure}
    \caption{Comparison of \ourSystem{} on image classification tasks.}
  \Description[]{}
  \label{fig_classificaiont}
\end{figure}

\textbf{Results.} 
Figure~\ref{fig_classificaiont} compares~\ourSystem{} with four baselines in terms of carbon emissions and top-1 inference accuracy. 
\ourSystem{} achieves 38.4\,gCO$_2$, reducing carbon emissions by~65.6\% relative to~RW's 111.7\,gCO$_2$ while preserving near-peak accuracy at~0.832 \vs~0.851. 
Compared with~DC and~EV, which both maintain full accuracy but incur~106.2 and~120.7\,gCO$_2$, respectively, \ourSystem{} reduces carbon emissions by~63.9\% and~68.2\%.
EE achieves the lowest carbon emissions at~16.4\,gCO$_2$ by selecting the lightest feasible models, but at a substantial accuracy cost of~0.759 \vs~0.832.
These results show that the benefits of~\ourSystem{} generalize across edge~AI services.

\subsubsection{Different Regions}

To evaluate generalizability under diverse grid conditions, we deploy all policies across six regions with markedly different carbon-intensity profiles. 
For example, PJM and ERCOT exhibit the highest mean carbon intensities, at~307.3~$\pm$~35.4 and~307.7~$\pm$~61.8\,gCO$_2$/kWh, respectively, whereas the hydropower-dominated BPAT grid is substantially cleaner at~37.8~$\pm$~17.1\,gCO$_2$/kWh.

\textbf{Results.}
Figure~\ref{fig_diff_area} shows that \ourSystem{} achieves consistent carbon reductions over the baselines in all six regions. 
Fossil-intensive regions such as~PJM and~ERCOT exhibit the highest emissions across all policies. 
In contrast, the hydropower-dominated~BPAT region remains below~20\,gCO$_2$ even for the least carbon-aware baselines, leaving limited room for further reduction. 
The benefit of~\ourSystem{} is most pronounced in high-carbon regions, where the~MPC planner shifts charging and battery use to cleaner time periods. 
Its gain is smallest in~BPAT, where the grid is already consistently low carbon. 
Across all regions, accuracy remains stable, ranging from~0.515 to~0.519~mAP. 
This shows that the carbon savings of~\ourSystem{} generalize across diverse grids without compromising inference quality.

\subsubsection{Different Battery Capacities}
We evaluate~\ourSystem{} under seven battery capacities of~5, 10, 18, 20, 30, 40, and~50\,Wh, while keeping all other parameters at their default settings.

\begin{figure}[t]
  \centering
  \begin{minipage}[b]{0.49\columnwidth}
    \centering
    \includegraphics[width=\linewidth]{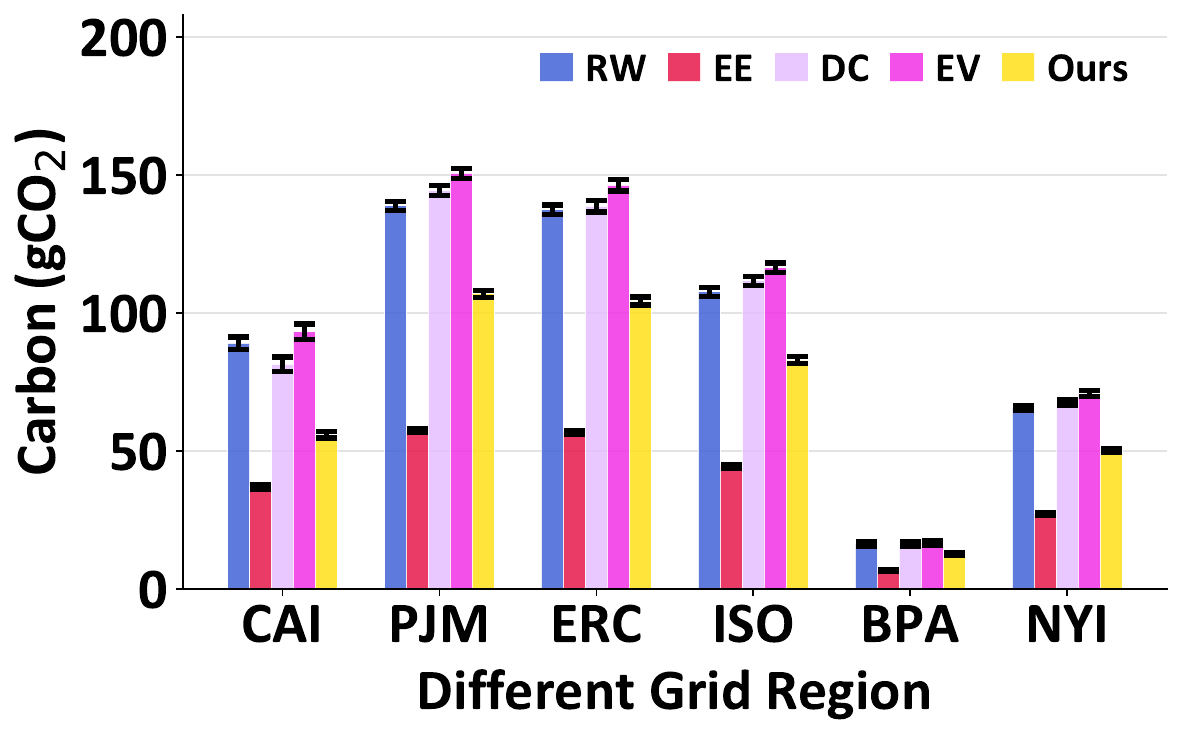} 
        \caption{Impact of region.}
    \Description[]{}
    \label{fig_diff_area}
  \end{minipage}
  \hfill
  \begin{minipage}[b]{0.49\columnwidth}
    \centering
    \includegraphics[width=\linewidth]{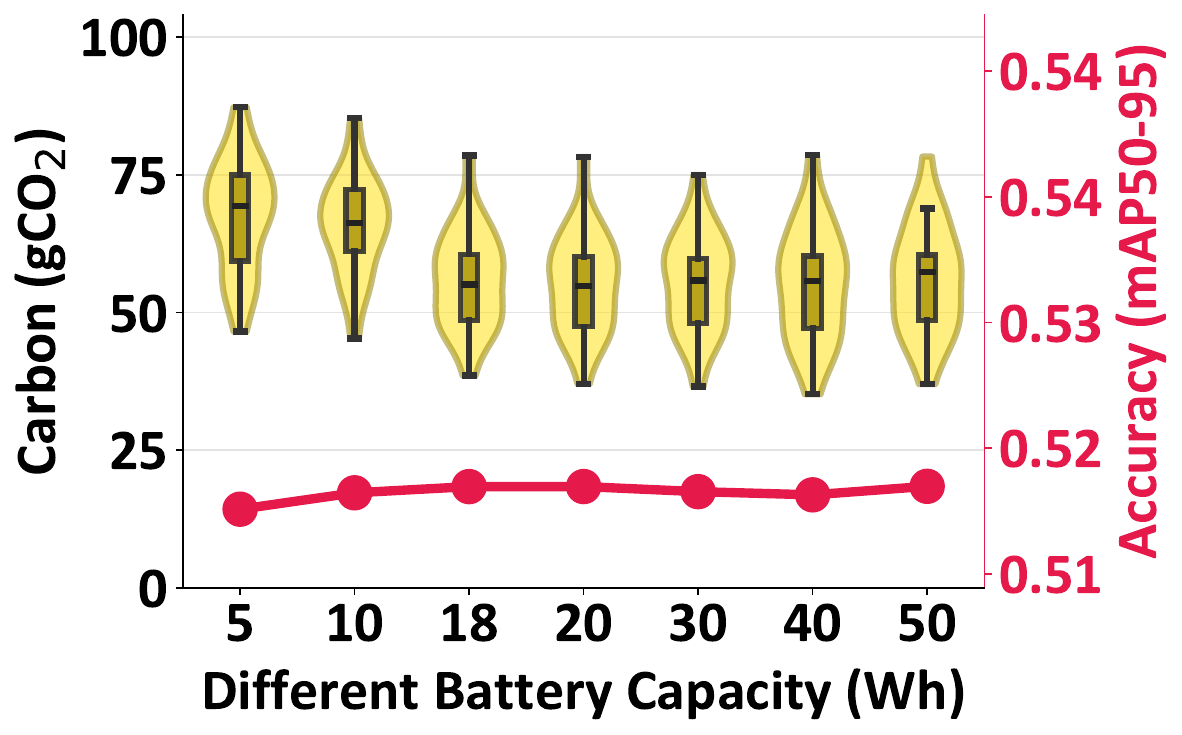}
        \caption{Impact of battery.}
    \Description[]{}
    \label{fig_diff_battery}
  \end{minipage}
\end{figure}

\textbf{Results.}
Figure~\ref{fig_diff_battery} shows that carbon emissions decrease as battery capacity increases from~5 to~18\,Wh, dropping from~67.8 to 55.9\,g\,CO$_2$, and then plateau over~20--50\,Wh at~55.0--55.8\,g\,CO$_2$. 
A smaller battery weakens the temporal energy buffer and forces more grid draws during high-carbon periods. 
Beyond~18\,Wh, additional capacity provides limited benefit, as the~MPC already has enough storage to shift charging into low-carbon windows. 
Despite a~23\% carbon gap between the smallest and largest batteries, accuracy remains stable at~0.516--0.518~mAP50. 
This indicates that~\ourSystem{} decouples inference quality from battery size by adapting charging schedules and model selection to the available energy buffer.

\begin{figure}[t]
  \centering
  \begin{minipage}[b]{0.49\columnwidth}
    \centering
    \includegraphics[width=\linewidth]{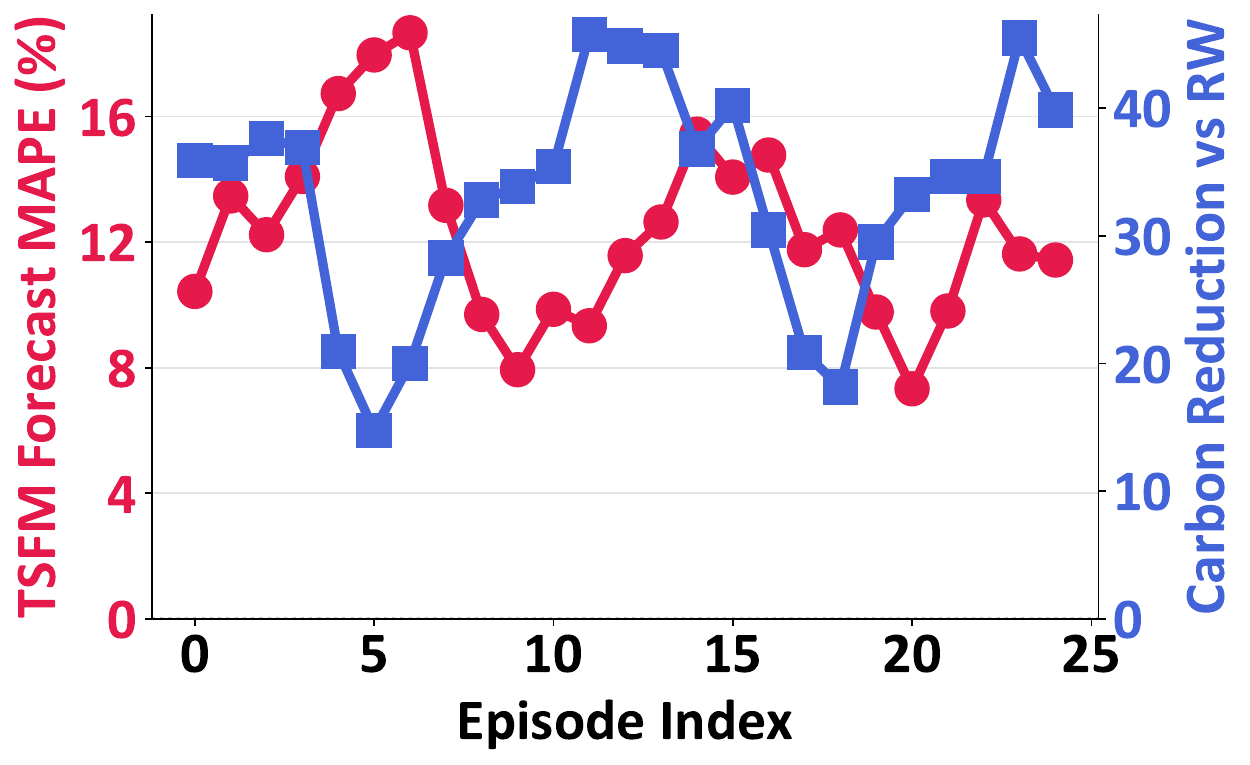} 
    \caption{TSFM forecast error.}
\Description[]{}
    \label{fig_tsfm}
  \end{minipage}
  \hfill
  \begin{minipage}[b]{0.49\columnwidth}
    \centering
    \includegraphics[width=\linewidth]{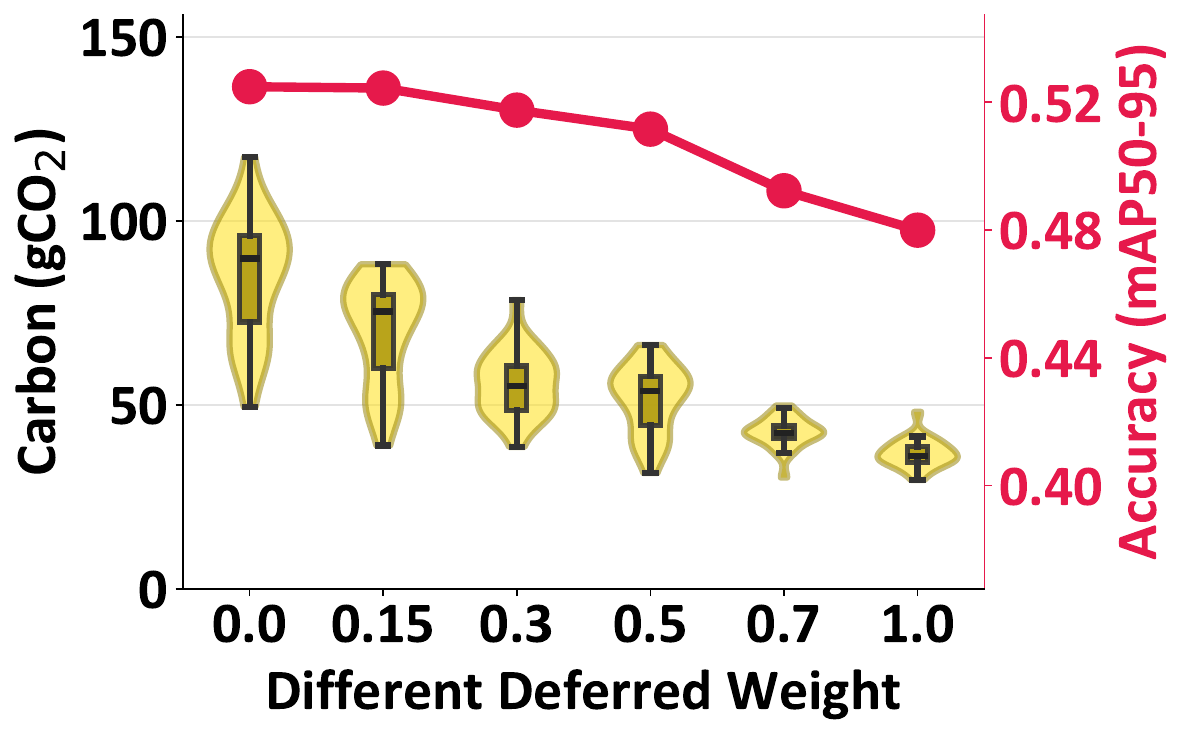}
    \caption{Deferred attribution.}
\Description[]{}
    \label{fig_defer}
  \end{minipage}
\end{figure}

\subsection{Ablation Study}
\label{sec:ablation}

\subsubsection{TSFM Forecasting Accuracy}
We analyze how forecasting accuracy affects the carbon reduction of \ourSystem{} relative to the RW baseline across~25 test episodes.
Figure~\ref{fig_tsfm} plots per-episode carbon intensity forecast mean absolute percentage error~(MAPE) on the left axis and the corresponding carbon reduction on the right axis.

\textbf{Results.}
Forecasting~MAPE ranges from~7.3\% to~18.7\%.
Despite this variation, \ourSystem{} achieves positive carbon reduction relative to~RW in every episode. 
The two curves exhibit a clear inverse relationship: lower forecast error generally corresponds to larger carbon reduction, whereas error spikes coincide with weaker savings. 
This indicates that improved forecasts enhance carbon-aware scheduling, while \ourSystem{} remains robust under degraded forecasts and degrades gracefully rather than failing abruptly.

\subsubsection{Deferred Attribution}

We evaluate six deferred-attribution weights:~0, 0.15, 0.3, 0.5, 0.7, and~1.0, where~0 treats battery discharge as carbon-free and~1.0 assigns the full estimated recharge cost.

\textbf{Results.}
Figure~\ref{fig_defer} shows that increasing the deferred-attribution weight from~0 to~1.0 monotonically reduces carbon emissions from 85.3 to~36.8\,g\,CO$_2$, as larger weights discourage battery discharge that would otherwise require recharge under high-carbon grid conditions. 
At zero weight, the free-discharge loophole causes the~DP solver to deplete the battery aggressively in support of heavier models, leading to higher recharge emissions later. 
In contrast, accuracy declines gradually from~0.525 to~0.480\,mAP, since larger weights bias the controller toward lighter and less accurate modes. 
The default weight of~0.3 provides a favorable trade-off, reducing carbon emissions by~34\% relative to the no-deferred-cost baseline while retaining~98.6\% of its inference accuracy.

\subsection{Overhead Study}
\label{sec:overhead}

\begin{table}[t]
\centering
\caption{Computational overhead on the Jetson Orin Nano.}
\label{tab:overhead}
\begin{tabular}{lll}
\toprule
\noindent\textbf{Component} & \noindent\textbf{Metric} & \noindent\textbf{Value (mean $\pm$ std)} \\
\midrule
TSFM Forecaster & Latency (ms)    & $333.3 \pm 314.1$ \\
                & GPU Memory (MB) & $593.7 \pm 0.0$ \\
\midrule
MPC Solver      & Latency (ms)    & $90.2 \pm 17.3$  \\
\bottomrule
\end{tabular}
\end{table}

Table~\ref{tab:overhead} summarizes the runtime overhead of~\ourSystem{} on the Jetson Orin Nano. 
The TSFM forecaster, Sundial-128M, incurs a first-call latency of~2024.7\,ms due to warm-up, while subsequent calls stabilize at~275.0\,ms, yielding a mean latency of~333.3\,ms~$\pm$~314.1\,ms. 
Since the~TSFM is invoked once every~$K = 96$ control slots, its amortized per-step overhead is below~3.5\,ms. 
The forecaster consumes~593.7\,MB of GPU memory, which fits within the Jetson's~8\,GB memory budget. 
The~DP solver adds~90.2~$\pm$~17.3\,ms per step. 
Overall, the combined per-step overhead remains well below the control interval of~$\Delta t = 15$\,min, demonstrating the practicality of~\ourSystem{} for real-time deployment on resource-constrained edge hardware.

\section{Conclusion}
\label{sec:conclusion}

We present~\ourSystem{}, a proactive carbon-aware adaptive control framework for edge~AI devices that treats the battery as an active temporal buffer. 
By combining zero-shot forecasting from edge-friendly~TSFMs with a receding-horizon dynamic programming solver, \ourSystem{} jointly optimizes the software pipeline variant, hardware operating point, and battery charging actions. 
Extensive experiments demonstrate the effectiveness of~\ourSystem{}.

\textbf{Discussion and Future Work.}
While~\ourSystem{} demonstrates promising carbon-reduction potential, several practical system challenges remain for future work.
\textit{First,} our current evaluation assumes a stable and continuous workload, whereas many edge applications, such as motion-triggered analytics, exhibit bursty and time-varying demand. 
Future work could incorporate stochastic workload prediction to enable preemptive energy buffering ahead of traffic surges.
\textit{Second,} the current controller does not account for mode-transition overheads. 
Incorporating explicit transition costs into the optimization could help prevent frequent mode switching.
\textit{Third,} \ourSystem{} relies on offline profiling to characterize the performance and energy profiles of different model and hardware configurations. 
To improve scalability, future work could replace this static profiling process with analytical or learned performance models.
\textit{Fourth,} our current design focuses on a single device with a paired battery. 
Future work could extend the framework to coordinated multi-device settings with a shared larger battery, \eg smart home scenarios.
\textit{Finally,} extending~\ourSystem{} to battery-only mobile devices, \eg smartphones, could require behavior-aware modeling of uncertain charging opportunities while preserving user-facing resilience.

\section*{Acknowledgments}
The research reported in this paper was sponsored in part
by the National Science Foundation through awards \# CNS-2211301, CNS-2211302, CNS-2213636, and CNS-2325956, and  by the DEVCOM ARL through award \# W911NF1720196. Any opinions, findings, and conclusions expressed in this material are those of the authors and do not necessarily reflect the views of the funding agencies.

\bibliographystyle{unsrt}

\end{document}